\documentclass{egpubl}
\usepackage{eg2025}
 
\ConferencePaper        
\CGFccbync

\usepackage[T1]{fontenc}
\usepackage{dfadobe}  

\usepackage{cite}  
\BibtexOrBiblatex
\electronicVersion
\PrintedOrElectronic
\ifpdf \usepackage[pdftex]{graphicx} \pdfcompresslevel=9
\else \usepackage[dvips]{graphicx} \fi

\usepackage{egweblnk} 

\usepackage{hyperref}
\usepackage{url}
\usepackage{wrapfig}
\usepackage{graphicx}
\usepackage{booktabs}
\usepackage{multirow}
\usepackage{gensymb}
\usepackage{layouts}
\usepackage{float}
\usepackage{amsmath}
\usepackage{amsfonts}

\title[Mesh Compression with Quantized Neural Displacement Fields]%
      {Mesh Compression with Quantized Neural Displacement Fields}

\author[Pentapati et al]
{\parbox{\textwidth}{\centering Sai Karthikey Pentapati$^{1,2}$\orcid{0009-0003-4763-8335}
, Gregoire Phillips$^2$, and Alan C. Bovik$^{1}$\orcid{0000-0001-6067-710X}
}
\\
{\parbox{\textwidth}{\centering $^1$Laboratory of Image and Video Engineering, The University of Texas at Austin\\
         $^2$Ericsson Inc., Santa Clara, CA
       }
}
}

\begin{document}

\teaser{
 \includegraphics[width=0.9\linewidth]{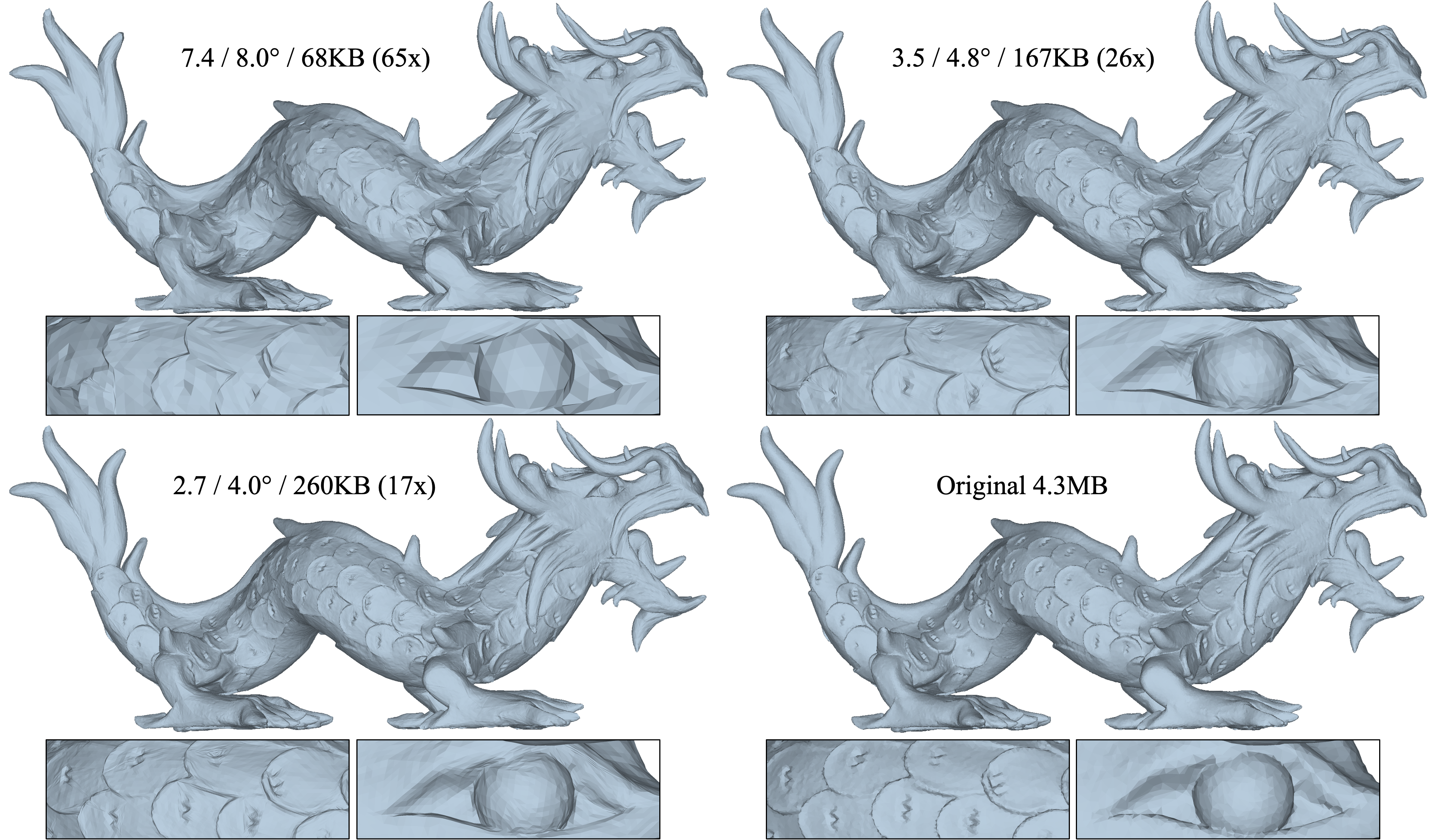}
 \centering
  \caption{We present a framework for compressing 3D meshes by compactly encoding a displacement field as an implicit neural representation. The displacement field can be applied to the surface of the coarse approximation of the original mesh to reconstruct it. The method achieves state-of-the-art reconstruction quality for a wide range of compression ratios. Here we show up to 65x compression of \texttt{XYZ} model courtesy of Stanford's 3D scanning repository. The point-to-mesh error, average normal error in degrees, and the compressed sizes are shown for each mesh.}
\label{fig:teaser}
}

\maketitle
\begin{abstract}
Implicit neural representations (INRs) have been successfully used to compress a variety of 3D surface representations such as Signed Distance Functions (SDFs), voxel grids, and also other forms of structured data such as images, videos, and audio. 
However, these methods have been limited in their application to unstructured data such as 3D meshes and point clouds. 
This work presents a simple yet effective method that extends the usage of INRs to compress 3D triangle meshes. 
Our method encodes a displacement field that refines the coarse version of the 3D mesh surface to be compressed using a small neural network. 
Once trained, the neural network weights occupy much lower memory than the displacement field or the original surface.
We show that our method is capable of preserving intricate geometric textures and demonstrates state-of-the-art performance for compression ratios ranging from 4x to 380x (See Figure \ref{fig:teaser} for an example). 


\end{abstract}  
\section{Introduction}
\label{sec:intro}
3D meshes, consisting of vertices, edges, and faces, are used to represent detailed geometric structures in computer graphics and are ubiquitously used in gaming, virtual reality, and simulation applications. 
The use of meshes has grown in popularity compared to other 3D representations because their efficient rendering has been facilitated by advancements in hardware, notably graphics processing units (GPUs), which accelerate the process through parallel computing and optimized graphics pipelines.
Despite these advancements in rendering technology, the inherent sizes and complexities of 3D meshes pose significant challenges in terms of memory use and transmission which are crucial for real-time communications applications.
The development of compression algorithms is therefore crucial to mitigate these challenges to ensure efficient handling and transmission of complex and large-scale 3D data.

In recent years, compression algorithms that rely on building Implicit Neural Representations (INRs) have been shown to outperform traditional methods across various modalities.
Among numerous 3D surface representations, the development of neural compression methods has been mostly focused on addressing the compression of Signed Distance Functions (SDFs) (\cite{Park_2019_CVPR, davies2021}) and occupancy grids (\cite{nglod, jiang2024cofie}).
However, the use of these representations is extremely limited in computer and mobile applications compared to meshes due to their inefficient rendering process given the available hardware.

Owing to these challenges, only a few deep compression techniques for meshes have been presented so far.
The primary challenge of formulating INR-based compression algorithms for meshes is their discrete and unstructured nature. 
It is not straightforward to encode the connections between vertices as neural representations, due to the challenge of posing the connectivity reconstruction problem as a gradient-based optimization problem.
To circumvent this challenge, remeshing and subdivision surfaces have been employed by \cite{neuralsubdivision} and \cite{npm} to reconstruct the vertex connectivity when building large datasets of simplified meshes and their high-polygon count versions of them.
Following a different direction, \cite{ngf} proposed a geometric primitive called the Neural Geometry Field to solve the problem of reconstructing the vertex connectivity.

In this paper, we present a simple method of constructing INRs of meshes and thereby compressing them.
Similar to previous works by \cite{npm} and\cite{neuralsubdivision}, our method also leverages remeshing and surface subdivisions to construct an as-structured-as-possible representation of a mesh to facilitate the training of an INR. 
Specifically, our method compresses a mesh by converting it to a very coarse approximation of its surface and a displacement field encoded as a highly compact INR that allows reconstruction of the original surface from the coarse approximation.
The major contributions of our work are the following:
\begin{enumerate}
    \item A method for generating a structured representation suitable for building an INR from a 3D mesh.
    \item Construction of a compact INR that very efficiently encodes a 3D surface represented as a mesh.
    \item Compression of the trained INR to obtain state-of-the-art mesh compression results
    \item The proposed overall scheme delivers very high compression outcomes which yield high-quality mesh reconstructions without significant increases in compute complexity.
\end{enumerate}

\section{Related Work}
\label{sec:related}
Our work furthers research on traditional and modern mesh compression algorithms, mesh surface subdivisions, and 3D surface representations using implicit neural networks.

Traditional methods of mesh compression can be broadly categorized into two types.
The first type includes methods that aim for lossless preservation of vertex connectivity. 
These mostly involve algorithms that traverse the vertices of a mesh with minimal repeated visits, which avoids duplicate vertex references when representing connectivity. These either use data structures like triangle fans/ triangle strips(\cite{ 10.1145/218380.218391, 10.5555/266989.267103, BAJAJ1999167}), spanning tree-based encoding (\cite{10.1145/274363.274365, dgpgmpr}), or triangle traversal encoding (\cite{edgebreaker, 10.1145/280814.280836, SZYMCZAK200153})
These methods are often paired with quantization and predictive coding of 3D coordinates of the vertices (\cite{10.1145/218380.218391, 10.1145/274363.274365, gotsman-touma-gi98}), and are now included in standard compression pipelines such as Google's Draco (\cite{draco}) which is used ubiquitously by the industry. 

The second type of classical methods are \textit{mesh simplification} methods, where the number of faces and vertices in the mesh are reduced while preserving as much geometric information as possible (\cite{qslim, qdssmp}).
However, The surface of a mesh can be \textit{subdivided} to increase its vertex resolution (\cite{ccsubdiv, sabin2002subdivision}.
In \cite{10.1145/237170.237216}, a method for progressively undoing the simplification is presented.
A simplified surface can also be subdivided to recover the original vertex resolution, albeit with limited fidelity of the reconstructed approximation.
Authors of \cite{10.1145/344779.344829} use a displacement map along with a subdivision surface to reconstruct a mesh and as such is very related to our work.
In their work, the displacement map is calculated as the signed distance between each vertex on the simplified and subdivided mesh, and the closest point on the surface of the original mesh. 
The major drawback is that if the simplified mesh has very low vertex resolution (which is desirable for compression), a greedy approach to locating the closest point on the original surface might yield unfavorable reconstructions.
The survey \cite{10.1145/2693443} contains more comprehensive descriptions of such mesh compression techniques.

\cite{neuralsubdivision} and \cite{npm} augment the reconstruction of original surfaces from their subdivided versions using learning-based approaches.
Both methods involve training large graph neural network-based deep-learning models that can upsample the vertex resolution of any given mesh by subdividing it, and predicting displacement offsets for each vertex based on their local geometry.
These methods use \textit{successive self-parameterization} (SSP) (\cite{neuralsubdivision} to build paired a dataset of vertex-aligned low and high polygon count meshes.
Their reliance on using only local geometry to compute displacement offsets and a single trained model to reconstruct all meshes limits the quality of their reconstructed outputs.
Instead of training one model to process all meshes, the methods devised by \cite{nvdiffmodelling} and \cite{ngf} perform per-mesh optimization. 
These methods rely on appearance-driven optimization of mesh simplification and building memory-efficient geometric fields, respectively.
Due to the high efficacy of per-mesh optimization, Neural Geometric Fields (NGF)(\cite{ngf}) achieves much better compression ratio-to-quality trade-offs than \cite{npm} and \cite {neuralsubdivision}.
Despite current state-of-the-art compression, appearance-driven optimization methods are prone to inaccurate reconstruction because they rely on minimizing the rendering loss.
The inaccuracies arise because minimizing the rendering loss to learn deformations to be applied to coarse surfaces as done in \cite{ngf} is akin to a greedy search for the closest points on the original surface to the points on the coarse surface, which might not be ideal. 

Mesh surfaces can also be cut along edge-paths and flattened and fit in a 2D square to obtain a "Geometry Image" (\cite{geoimg, 10.1007/3-540-26808-1_2}). 
This "Geometry Image" can be compressed using any popular 2D image compression methods, such as wavelet-based image encoders.
Authors of \cite{morreale2022neural} present a method to build compressed neural representations of Geometry Images of 3D surfaces by fitting surface-wise convolutional neural networks.

Per-surface optimization methods for building INRs and geometric primitives have also been successfully applied to compress SDFs (\cite{sitzmann2019siren, Park_2019_CVPR, taast, yenamandra2024fire})  and occupancy grids (\cite{tancik2020fourfeat, LIU2021107865, tang2021octfield, nglod, acorn}).
Many works also leverage hybrid representations.
For example, \cite{jiang2024cofie}  builds a generalized surface representation by introducing Coordinate Fields that are hierarchical voxel grids with latent codes per each cell that allow decoding the SDF.
\cite{sdmnet} partition meshes into bounding boxes and use a network to apply deformations on them.
INR-based image  (\cite{strumpler2022inrcompress}) and video (\cite{chen2021nerv}) compression methods have also been shown to outperform traditional methods.
Quantizing the parameters of INRs has often led to significantly increased compression ratios in many works of this nature. 
For instance, \cite{vbrt} compress implicit feature grids that encode SDFs using vector-quantized dictionaries, \cite{ntc2023} build compressed representations of mesh texture by quantizing neural features, while \cite{zhang2024rate, Gordon_2023_WACV, kang2022ternarynerf} quantize the parameters of a neural radiance field (NeRF) \cite{nerf} to build compact reconstructions of 3D scenes.

Inspired by the success of per-mesh and per-surface optimization methods and INRs for compression, we propose a method of mesh compression by encoding the displacement field as a compact neural network that is trained specifically for each mesh. 
The displacement field can refine the simplified version of the coarse mesh and build an accurate reconstruction of the original surface.

\section{Neural Mesh Compression}
\label{sec:method}

\subsection{Prerequisite: Successive Self-Parameterization}
\label{subsec:ssp}
\begin{figure}
  \begin{center}
    \includegraphics[width=0.40\textwidth]{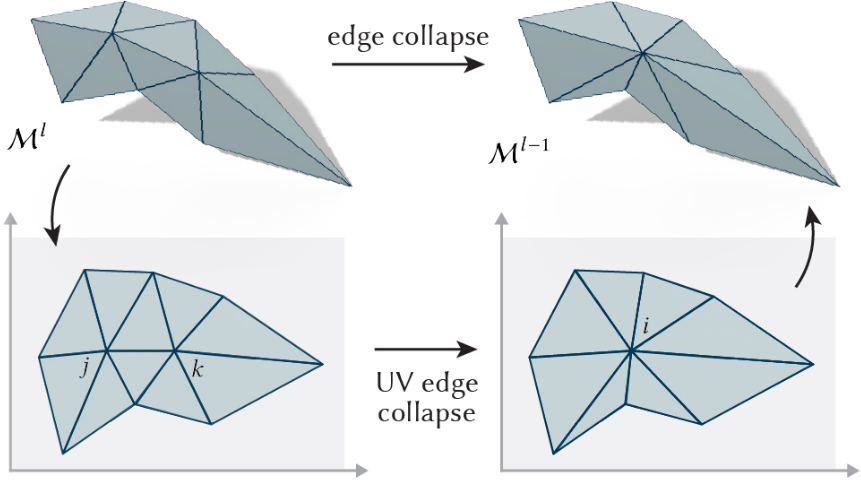}
  \end{center}
  \caption{Successive self-parameterization. (Figure borrowed from \cite{neuralsubdivision})}
  \label{fig:ssp}
\end{figure}

Successive self-parametrization, proposed by \cite{neuralsubdivision}, is a method for obtaining a bijective mapping between the surface of a mesh and its simplified representation.
We employ this technique in our method for remeshing the surface to be compressed and thus include this subsection for readers' ease of understanding.
SSP takes a triangle mesh and an edge-collapse algorithm as input, then generates a decimated triangle mesh along with a bijective map between the surfaces of the original and decimated meshes.
This is achieved by simultaneous collapse operations in the 3D domain and the UV domain consisting of the flattened 1-ring of the edge as shown in Figure \ref{fig:ssp} (borrowed from \cite{neuralsubdivision}).
Due to the imposed boundary constraint in the UV domain, there is a bijective mapping between the UV patches before and after the edge collapse.
The bijective map in the UV domain also allows a bijective mapping between the pre- and post-collapse surfaces in the 3D domain.

\subsection{Method Overview}
\begin{figure*}[t]
    \centering
    \includegraphics[width=0.95\linewidth]{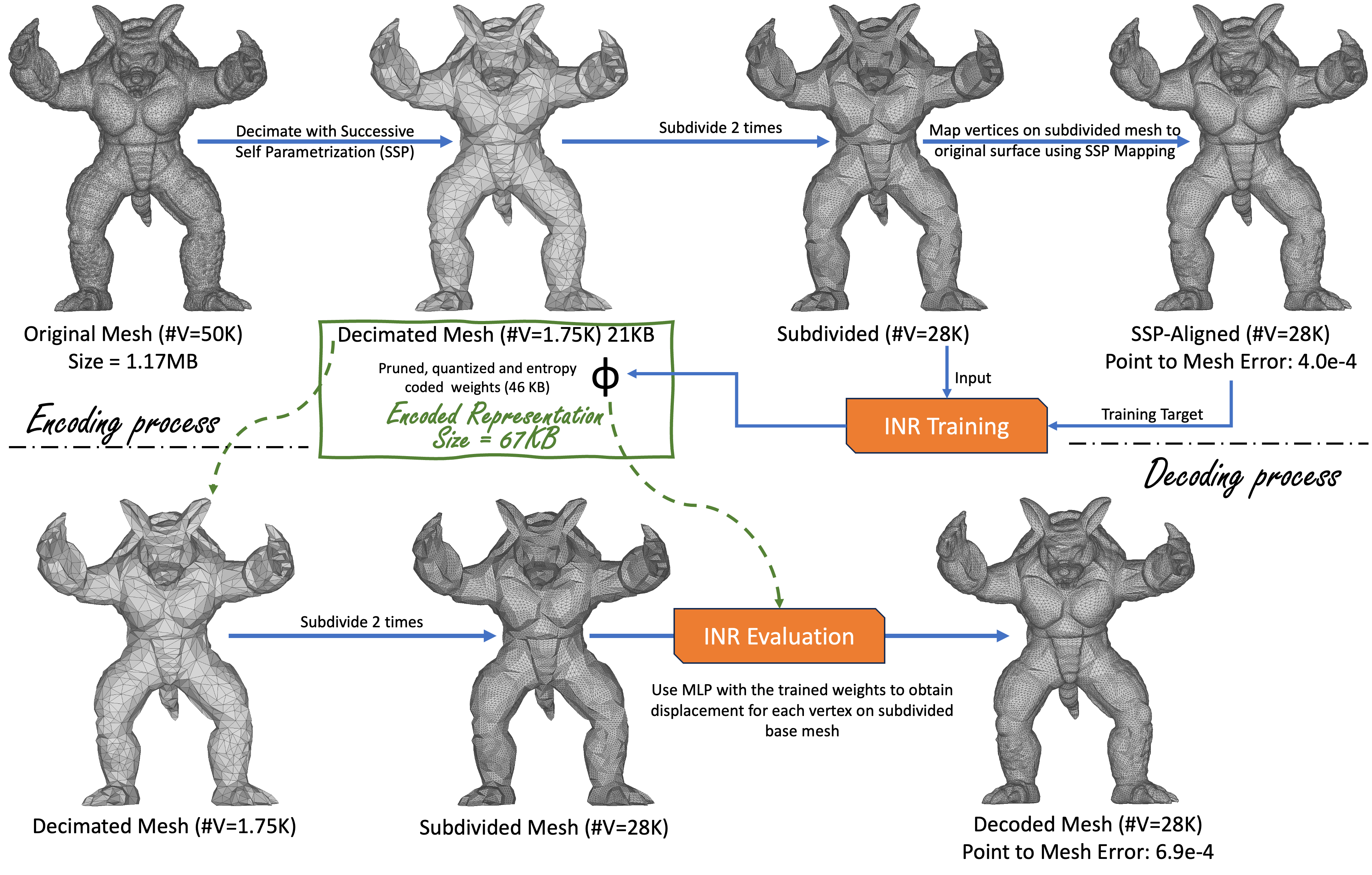}
    \caption{Overview of Neural Mesh Compression}
    \label{fig:overview}
\end{figure*}

Similar to methods like \cite{nglod} and \cite{ngf}, we train a neural network model specifically for each mesh to be compressed.
Our method for building compressed representations of meshes can broken down into the following steps and is also illustrated in Figure \ref{fig:overview}:
\begin{enumerate}
    \item Decimate the mesh to be compressed by reducing the number of faces by a large extent to obtain its coarse version, while simultaneously performing successive self-parametrization (see section \ref{subsec:ssp}) to obtain a bijective mapping between surfaces of the original and the decimated meshes.
    \item Perform midpoint subdivision of the faces of the coarse mesh a desired number of times to increase the sampling resolution of the surface of the decimated mesh.
    \item Use the bijective map generated by SSP to determine a displacement map for the vertices of the subdivided mesh, such that the displaced vertices reconstruct the original surface.
    \item Overfit a simple multi-layer perceptron (MLP) that encodes the displacement of each vertex of the super-sampled subdivided mesh.
    \item Prune, quantize, and entropy-encode the weights of the trained model.
\end{enumerate}

The compressed representation of the mesh now consists of the coded weights of the overfitted network and the coarse mesh.
To decode and reconstruct the mesh, the compression can be reversed according to the following steps:
\begin{enumerate}
    \item Decode the entropy-coded quantized weights and load them to the model.
    \item Perform midpoint subdivision of the coarse mesh the same number of times as in the encoding process.
    \item Use the quantized model to compute the displacement that needs to be applied to each vertex on the subdivided mesh.
    \item Apply the computed displacement to obtain a decoded approximation of the original surface.
\end{enumerate}

We explain the design of the encoding process of Neural Mesh Compression in the subsequent subsections.

\subsection{Generating Training Data}
\label{subsec:datasest}
As our method involves training a neural network specific to each mesh, a training dataset with paired inputs and targets has to be generated for each mesh to be compressed.
Given a mesh $\text{M}_\text{orig}:=(\text{V}_\text{orig}, \text{F}_\text{orig})$ to be compressed, we define its surface ($\text{S}_\text{orig} \subset \mathbb{R}^3$) as the set of its vertices, all points lying on its edges, and all points lying within its faces. 
First, $\text{M}_\text{orig}$ is simplified using \texttt{QSLIM} (\cite{qslim}) to obtain $\text{M}_\text{coarse}$, which is a very coarse approximation of the original, such that $|\text{V}_\text{orig}| \gg |\text{V}_\text{coarse}|$. 
The simplification process is performed with SSP to obtain a bijective function $f_\text{SSP}: \text{S}_\text{coarse} \to \text{S}_\text{orig}$.
Thus, the original surface can be reconstructed by transforming each point on the coarse surface using the mapping. This process can be formulated as $\text{S}_\text{orig}=\{f_\text{SSP}(p) | p \in \text{S}_\text{coarse}\}$.
Then, a displacement field, $\mathcal{F}:\text{S}_\text{coarse}\to\mathbb{R}^3$ can be defined such that $\mathcal{F}(p) = f_\text{SSP}(p)-p, \forall p\in\text{S}_\text{coarse}$.
With this proposed formulation, the task of mesh compression can be converted into the task of compressing the field $\mathcal{F}$ that is defined on a topological surface $\text{S}_\text{coarse}$.
INRs have been shown to be amenable to compressing such fields in works by \cite{sitzmann2019siren} and \cite{yifan2022geometryconsistent}. 

So far, with this method of point-wise surface mapping, infinite pairs of points and their corresponding displacements can be sampled, yielding data that can be used to train an INR that perfectly captures the geometry of the surface.
This, however, still leaves the task of compressing the vertex connectivity of the mesh.
When conducting reconstruction and decoding, the point cloud obtained by evaluating the INR can be triangulated to obtain a triangle mesh, but errors or inaccuracies in the trained INR can lead to poor triangulation.
Also, an INR training pipeline with randomly sampled pairs in each batch tends to slow the training down significantly due to the high time cost of evaluating $\mathcal{F}(p)$ relative to performing gradient descent on a small MLP.

Pre-defining a subdivision scheme, and hence also a surface sampling scheme, addresses both of these issues.
We employ a midpoint subdivision scheme to increase the vertex resolution of the coarse surface, whereby a new vertex is introduced at the midpoint of each edge of the mesh.
We subdivide $\text{M}_\text{coarse}$ a fixed number of times (say $s$ times) to obtain $\text{M}_\text{subdivided}$, such that $|\text{F}_\text{subdivided}| = 4^s\times|\text{F}_\text{coarse}|$.
Note that this subdivision scheme ensures that the same vertex connectivity is maintained during the encoding and decoding process.
This implies that no additional encoding is required for reconstructing connectivity.
Additionally, because the vertex coordinates are the same during the encoding and decoding processes, the vertices introduced by successive subdivisions are also exactly the same. 
This implies that the inputs to the INR during training and decoding can be predetermined. 
The predetermination of inputs allows caching the training dataset, since the paired set $\{(\mathcal{F}(p), p): \forall p \in \text{V}_\text{subdivided}]\}$ need only be computed once, and each sample can be used across multiple training batches.
It is crucial to note that finite sampling of $\text{S}_\text{coarse}$ implies that only an approximation of $\text{S}_\text{orig}$ is used as the training target for the INR. 
In section \ref{subsec:ablation}, we find error bounds enforced by this approximation.

A neural network can now be fitted to this dataset to obtain an INR of the displacement field needed to reconstruct an accurate approximation of $\text{S}_\text{orig}$ from $\text{S}_\text{subdivided}$.

\subsection{Neural Mesh Compression Architecture}
As deep neural networks are great universal function approximations, we employ a multi-layer perception (MLP)  with parameters $\theta$ to approximate the displacement field $\mathcal{F}$ to be applied on $\text{V}_\text{subdivided}$. 
The proposed architecture for mesh compression consists of generating positional embedding of inputs fed to a dense multi-layer perceptron in which the first few layers are augmented with newly proposed one-ring feature aggregation. 

Instead of feeding the 3D coordinates of vertices $p \in \text{V}_\text{subdivided}$ (where $\text{V}_\text{subdivided} \subset \mathbb{R}^3$) to the network directly, we first perform \textbf{positional encoding} to embed them in a higher dimensional frequency space, similar to \cite{nerf}.
Specifically, we map each real-valued coordinate from $\mathbb{R}$ to the frequency space according to Equation \ref{eq:pe} to get a vector in $\mathbb{R}^\text{2Q}$, where $Q$ is a hyperparameter.
As this transformation is performed on each of the three vertex coordinates, the coordinates $\text{V}_\text{subdivided} \subset \mathbb{R}^3$ get mapped to vectors having 6Q dimensions, which are then fed to the MLP.
This addresses the issue of neural networks being biased towards learning low-frequency functions that was studied by (\cite{Rahaman2018OnTS}).

\begin{multline}
    \gamma(p) = ( \sin(\pi p \cdot 2^0), \cos(\pi p \cdot 2^0), \ldots, \\\sin(\pi p \cdot 2^{Q-1}), \cos(\pi p \cdot 2^{Q-1}) )
\label{eq:pe}
\end{multline}

\begin{figure}
  \begin{center}
    \includegraphics[width=0.4\textwidth]{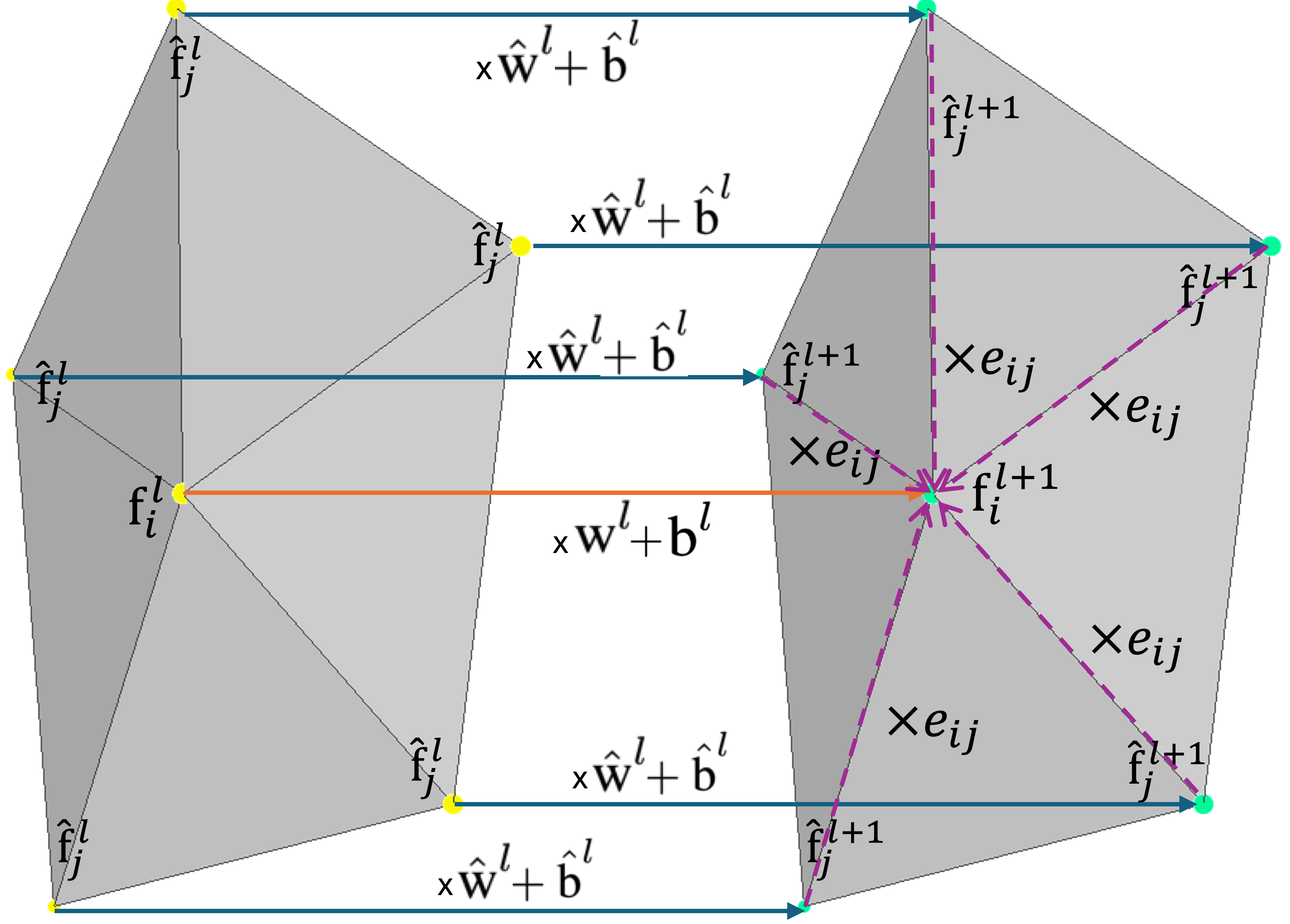}
  \end{center}
  \label{fig:onering}
  \caption{One-ring feature accumulation}
\end{figure}

Additionally, we observe that having access to the local geometry of a vertex improves the ability of the neural network to encode $\mathcal{F}$.
To that end, we devised \textbf{one-ring feature accumulation}.
Given a vertex  $i$ being processed, its activations at layer $l$ ($\text{f}^l_i$) are aggregated with the activations of the vertices in its one-ring neighborhood ($\hat{\text{f}}^l_j\forall j \in \mathcal{N}(i)$) as shown in Equation \ref{eq:onering}, where $\text{w}^l$ and $\text{b}^l$ are the weights and bias of the $l^{th}$ layer of the main branch of the MLP, and $\hat{\text{w}}^l$, $\hat{\text{b}}^l$ are the weights and biases of the $l^{th}$ auxiliary branches of the MLP responsible for computing the neighborhood features, and $e_{ij}$ is the length of the edge connecting vertices $i$ and $j$.
This operation is also visualized in the Figure \ref{fig:onering}. 
This operation is a more localized and runtime-memory-efficient variant of graph convolutions.
Performing this operation instead of regular graph convolutions reduces the time and GPU memory required to train the network drastically, while yielding similar reconstruction quality (See section \ref{subsec:ablation}). 
Only the first few layers of the multi-layer perceptron are augmented with one-ring feature accumulation.
The feature accumulation process is:
\begin{equation}
\begin{aligned}
    \hat{\text{f}}^{l+1}_j &= \hat{\text{w}}^l\cdot \hat{\text{f}}^l_j + \hat{\text{b}}^l, \forall j\in\mathcal{N}(i) \\
    \text{f}^{l+1}_i &= (\text{w}^l\cdot \text{f}^l_i + \text{b}^l + \sum_{j\in\mathcal{N}(i)} (\hat{\text{f}}^{l+1}_j\cdot e_{ij})) \times 0.5
\label{eq:onering}
\end{aligned}
\end{equation}
The difference between one-ring accumulation and graph convolution operations \cite{kipf2017semi, milano2020primal} is subtle yet significant.
In a graph neural network built with $N$ graph convolution layers in series, any vertex $i$ will be influenced by all the vertices that are at most $N$ edges away from $i$. 
However, in the case of an $N$-layered MLP with one-ring accumulation, any vertex $i$ is only influenced by vertices that are at most one (and not $N$) edges away.
This is because the features $\hat{\text{f}}$ of each vertex are obtained from an independent MLP before being aggregated with features "$\text{f}$" of their one-ring neighbors that are extracted from the main MLP, as described in Equation \ref{eq:onering}.

We advise against using batch normalization in the neural network.
Since building an INR involves overfitting on the training dataset, using batch normalization modifies the computed intermediate features based on the other values in the training mini-batch.
This creates an effect of pseudo-randomization of the features of a sample during training, due to the impact of other samples in its batch which hinders the overfitting process.
In our implementation, we use layer-normalization to ensure deterministic training.

The overall architecture of the INR is quite simple. 
It is a multi-layer perceptron with $l$ hidden layers and $k$ features. Among the $h$ hidden layers, the first $g$ layers are augmented with one-ring feature accumulation.
Section \ref{subsec:setup} provides the exact values of these and other hyper-parameters.

\subsection{Compressing The INR Parameters}
\label{subsec:modelcomp}
We employ pruning, quantization, and entropy coding to further compress the parameters $\theta$ of the MLP that encodes the displacement field.

\textbf{L1 unstructured pruning} zeroes those parameters with the lowest $(1-S)\times 100$ percentile of L1-norm, to obtain a sparsity factor of $S$ (the fraction of non-zero parameters).
Instead of pruning the model to obtain a sparsity factor of $S$ in one shot after training, we prune it progressively, which involves pruning the model in multiple steps during the course of model training.
To reach a target sparsity of $S\%$, the model can be pruned and fine-tuned alternatively $z$ times where each iteration of pruning introduces sparsity by a factor of $s = \sqrt[z]{S}$.

All the parameters of the INR that were originally represented in a 32-bit precision floating point format, are quantized to 8-bit integers.
We employ a simple \textbf{post-training quantization} strategy, where all parameters are first normalized to $[0,1]$, scaled by 256, and rounded to the nearest integer. 
However, we dynamically quantize the activations based on their range during inference to improve accuracy.

One must note that, while pruning may suppress a parameter to zero, it still occupies 32 bits (or 8 bits after quantization) of storage.
To reap the benefits of improved compression, the network must be pruned to a high factor ($\leq0.6$ for instance), leading to a high frequency of zeroes in the list of parameters. 
\textbf{Entropy coding} can then yield a significantly more compact representation.
To this end, we apply Huffman Coding on the quantized weights, to obtain a bit-stream that can be decoded losslessly.

\section{Experiments \& Results}
\label{sec:results}

\newcommand{\customsize}{\fontsize{6.6}{10}\selectfont}
\setlength{\tabcolsep}{4.3pt}
\renewcommand{\arraystretch}{1.4}
\begin{table*}[ht]
\centering
\customsize
\begin{tabular}{|c|cccc|cccc|cccc|cccc|}
\hline
\textbf{Compressed Sizes} &
  \multicolumn{4}{c|}{\textbf{85KB}} &
  \multicolumn{4}{c|}{\textbf{130KB}} &
  \multicolumn{4}{c|}{\textbf{187KB}} &
  \multicolumn{4}{c|}{\textbf{260KB}} \\ \hline
\textbf{Mesh (Size MB)} &
  \multicolumn{1}{c|}{\textbf{Ours}} &
  \multicolumn{1}{c|}{\textbf{\begin{tabular}[c]{@{}c@{}}Ours \\ NPQC\end{tabular}}} &
  \multicolumn{1}{c|}{\textbf{NGF}} &
  \textbf{\begin{tabular}[c]{@{}c@{}}QS-\\ DRC\end{tabular}} &
  \multicolumn{1}{c|}{\textbf{Ours}} &
  \multicolumn{1}{c|}{\textbf{\begin{tabular}[c]{@{}c@{}}Ours\\ NPQC\end{tabular}}} &
  \multicolumn{1}{c|}{\textbf{NGF}} &
  \textbf{\begin{tabular}[c]{@{}c@{}}QS-\\ DRC\end{tabular}} &
  \multicolumn{1}{c|}{\textbf{Ours}} &
  \multicolumn{1}{c|}{\textbf{\begin{tabular}[c]{@{}c@{}}Ours\\ NPQC\end{tabular}}} &
  \multicolumn{1}{c|}{\textbf{NGF}} &
  \textbf{\begin{tabular}[c]{@{}c@{}}QS-\\ DRC\end{tabular}} &
  \multicolumn{1}{c|}{\textbf{Ours}} &
  \multicolumn{1}{c|}{\textbf{\begin{tabular}[c]{@{}c@{}}Ours\\ NPQC\end{tabular}}} &
  \multicolumn{1}{c|}{\textbf{NGF}} &
  \textbf{\begin{tabular}[c]{@{}c@{}}QS-\\ DRC\end{tabular}} \\ \hline
\textbf{Armadillo} (1.14) &
  \multicolumn{1}{c|}{\textbf{6.92}} &
  \multicolumn{1}{c|}{12.03} &
  \multicolumn{1}{c|}{13.54} &
  8.71 &
  \multicolumn{1}{c|}{\textbf{4.91}} &
  \multicolumn{1}{c|}{9.55} &
  \multicolumn{1}{c|}{9.80} &
  6.07 &
  \multicolumn{1}{c|}{\textbf{3.46}} &
  \multicolumn{1}{c|}{7.25} &
  \multicolumn{1}{c|}{8.42} &
  5.01 &
  \multicolumn{1}{c|}{2.70} &
  \multicolumn{1}{c|}{5.53} &
  \multicolumn{1}{c|}{7.67} &
  \textbf{2.55} \\
\textbf{Dragon} (12.51) &
  \multicolumn{1}{c|}{\textbf{18.15}} &
  \multicolumn{1}{c|}{26.44} &
  \multicolumn{1}{c|}{26.66} &
  29.11 &
  \multicolumn{1}{c|}{\textbf{11.24}} &
  \multicolumn{1}{c|}{20.72} &
  \multicolumn{1}{c|}{18.60} &
  16.70 &
  \multicolumn{1}{c|}{\textbf{8.51}} &
  \multicolumn{1}{c|}{15.33} &
  \multicolumn{1}{c|}{14.97} &
  11.83 &
  \multicolumn{1}{c|}{\textbf{6.46}} &
  \multicolumn{1}{c|}{13.88} &
  \multicolumn{1}{c|}{13.37} &
  7.50 \\
\textbf{Einstein} (8.44) &
  \multicolumn{1}{c|}{\textbf{7.90}} &
  \multicolumn{1}{c|}{14.42} &
  \multicolumn{1}{c|}{17.90} &
  19.12 &
  \multicolumn{1}{c|}{\textbf{5.23}} &
  \multicolumn{1}{c|}{9.89} &
  \multicolumn{1}{c|}{8.90} &
  9.29 &
  \multicolumn{1}{c|}{\textbf{3.90}} &
  \multicolumn{1}{c|}{8.04} &
  \multicolumn{1}{c|}{8.05} &
  5.53 &
  \multicolumn{1}{c|}{\textbf{2.63}} &
  \multicolumn{1}{c|}{6.11} &
  \multicolumn{1}{c|}{5.41} &
  3.12 \\
\textbf{Ganesha} (31.61) &
  \multicolumn{1}{c|}{\textbf{7.50}} &
  \multicolumn{1}{c|}{14.06} &
  \multicolumn{1}{c|}{13.75} &
  16.61 &
  \multicolumn{1}{c|}{\textbf{4.80}} &
  \multicolumn{1}{c|}{9.28} &
  \multicolumn{1}{c|}{9.89} &
  10.77 &
  \multicolumn{1}{c|}{\textbf{3.87}} &
  \multicolumn{1}{c|}{8.46} &
  \multicolumn{1}{c|}{8.66} &
  7.38 &
  \multicolumn{1}{c|}{\textbf{2.89}} &
  \multicolumn{1}{c|}{6.42} &
  \multicolumn{1}{c|}{5.81} &
  6.03 \\ 
\textbf{Gargoyle} (2.32) &
  \multicolumn{1}{c|}{\textbf{4.38}} &
  \multicolumn{1}{c|}{9.30} &
  \multicolumn{1}{c|}{12.27} &
  16.36 &
  \multicolumn{1}{c|}{\textbf{2.85}} &
  \multicolumn{1}{c|}{6.87} &
  \multicolumn{1}{c|}{8.70} &
  7.35 &
  \multicolumn{1}{c|}{\textbf{2.33}} &
  \multicolumn{1}{c|}{4.97} &
  \multicolumn{1}{c|}{6.69} &
  5.62 &
  \multicolumn{1}{c|}{\textbf{1.69}} &
  \multicolumn{1}{c|}{3.76} &
  \multicolumn{1}{c|}{5.11} &
  2.07 \\ 
\textbf{Gnome} (2.25) &
  \multicolumn{1}{c|}{\textbf{7.22}} &
  \multicolumn{1}{c|}{10.61} &
  \multicolumn{1}{c|}{8.43} &
  12.93 &
  \multicolumn{1}{c|}{\textbf{3.73}} &
  \multicolumn{1}{c|}{7.75} &
  \multicolumn{1}{c|}{6.49} &
  8.33 &
  \multicolumn{1}{c|}{\textbf{2.16}} &
  \multicolumn{1}{c|}{4.98} &
  \multicolumn{1}{c|}{4.97} &
  5.94 &
  \multicolumn{1}{c|}{\textbf{1.65}} &
  \multicolumn{1}{c|}{3.52} &
  \multicolumn{1}{c|}{3.24} &
  2.25 \\ 
\textbf{Head} (4.23) &
  \multicolumn{1}{c|}{\textbf{3.61}} &
  \multicolumn{1}{c|}{6.75} &
  \multicolumn{1}{c|}{10.04} &
  13.61 &
  \multicolumn{1}{c|}{\textbf{2.37}} &
  \multicolumn{1}{c|}{5.00} &
  \multicolumn{1}{c|}{6.16} &
  8.23 &
  \multicolumn{1}{c|}{\textbf{1.82}} &
  \multicolumn{1}{c|}{3.67} &
  \multicolumn{1}{c|}{5.32} &
  4.36 &
  \multicolumn{1}{c|}{\textbf{1.43}} &
  \multicolumn{1}{c|}{2.82} &
  \multicolumn{1}{c|}{4.57} &
  1.67 \\ 
\textbf{Lucy} (6.08) &
  \multicolumn{1}{c|}{\textbf{5.22}} &
  \multicolumn{1}{c|}{9.69} &
  \multicolumn{1}{c|}{12.94} &
  21.69 &
  \multicolumn{1}{c|}{\textbf{3.16}} &
  \multicolumn{1}{c|}{7.16} &
  \multicolumn{1}{c|}{10.57} &
  8.74 &
  \multicolumn{1}{c|}{\textbf{2.40}} &
  \multicolumn{1}{c|}{5.23} &
  \multicolumn{1}{c|}{4.54} &
  3.99 &
  \multicolumn{1}{c|}{\textbf{1.79}} &
  \multicolumn{1}{c|}{3.94} &
  \multicolumn{1}{c|}{4.28} &
  2.88 \\ 
\textbf{Metatron} (1.63) &
  \multicolumn{1}{c|}{\textbf{3.90}} &
  \multicolumn{1}{c|}{7.39} &
  \multicolumn{1}{c|}{14.65} &
  8.95 &
  \multicolumn{1}{c|}{\textbf{2.71}} &
  \multicolumn{1}{c|}{5.35} &
  \multicolumn{1}{c|}{12.36} &
  6.15 &
  \multicolumn{1}{c|}{\textbf{2.04}} &
  \multicolumn{1}{c|}{3.91} &
  \multicolumn{1}{c|}{11.65} &
  3.17 &
  \multicolumn{1}{c|}{\textbf{1.53}} &
  \multicolumn{1}{c|}{2.89} &
  \multicolumn{1}{c|}{9.53} &
  1.74 \\
\textbf{XYZ} (2.95) &
  \multicolumn{1}{c|}{\textbf{6.90}} &
  \multicolumn{1}{c|}{12.92} &
  \multicolumn{1}{c|}{14.41} &
  22.60 &
  \multicolumn{1}{c|}{\textbf{4.61}} &
  \multicolumn{1}{c|}{8.34} &
  \multicolumn{1}{c|}{12.64} &
  9.63 &
  \multicolumn{1}{c|}{\textbf{3.36}} &
  \multicolumn{1}{c|}{6.14} &
  \multicolumn{1}{c|}{8.04} &
  6.84 &
  \multicolumn{1}{c|}{\textbf{2.70}} &
  \multicolumn{1}{c|}{5.37} &
  \multicolumn{1}{c|}{6.79} &
  3.98 \\ \hline
\textbf{Mean Error} &
  \multicolumn{1}{c|}{\textbf{7.17}} &
  \multicolumn{1}{c|}{12.36} &
  \multicolumn{1}{c|}{14.46} &
  16.97 &
  \multicolumn{1}{c|}{\textbf{4.56}} &
  \multicolumn{1}{c|}{8.99} &
  \multicolumn{1}{c|}{10.41} &
  9.12 &
  \multicolumn{1}{c|}{\textbf{3.39}} &
  \multicolumn{1}{c|}{6.80} &
  \multicolumn{1}{c|}{8.13} &
  5.97 &
  \multicolumn{1}{c|}{\textbf{2.55}} &
  \multicolumn{1}{c|}{5.42} &
  \multicolumn{1}{c|}{6.58} &
  3.38 \\ \hline
\end{tabular}%
\caption{Each mesh is first compressed to 4 bitrates using our method, our method without pruning, quantization, and entropy coding of parameters (Ours NPQC), NGF, and QS-DRC. Each cell denotes $d_{pm}$ scaled $\times 10^4$ evaluated between the original mesh and the reconstructed meshes. The number of faces in the selected meshes varies from $50K$ in \textbf{Armadillo} to 2.1M in \textbf{Ganesha}. \textbf{Armadillo}, \textbf{Lucy}, and \textbf{XYZ} are courtesy of Stanford's 3D scanning repository; \textbf{Ganesha} is courtesy of \textit{Peel3D}, while \textbf{Dragon}, \textbf{Einstein}, \textbf{Metatron}, and \textbf{Gargoyle} are available in \textit{Thingi10K} (\cite{Thingi10K}).}
\label{tab:results-dpm}
\end{table*}

\begin{table*}[ht]
\centering
\customsize
\begin{tabular}{|c|cccc|cccc|cccc|cccc|}
\hline
\textbf{Compressed Sizes} &
  \multicolumn{4}{c|}{\textbf{85KB}} &
  \multicolumn{4}{c|}{\textbf{130KB}} &
  \multicolumn{4}{c|}{\textbf{187KB}} &
  \multicolumn{4}{c|}{\textbf{260KB}} \\ \hline
\textbf{Mesh (Size MB)} &
  \multicolumn{1}{c|}{\textbf{Ours}} &
  \multicolumn{1}{c|}{\textbf{\begin{tabular}[c]{@{}c@{}}Ours\\ NPQC\end{tabular}}} &
  \multicolumn{1}{c|}{\textbf{NGF}} &
  \textbf{\begin{tabular}[c]{@{}c@{}}QS-\\ DRC\end{tabular}} &
  \multicolumn{1}{c|}{\textbf{Ours}} &
  \multicolumn{1}{c|}{\textbf{\begin{tabular}[c]{@{}c@{}}Ours\\ NPQC\end{tabular}}} &
  \multicolumn{1}{c|}{\textbf{NGF}} &
  \textbf{\begin{tabular}[c]{@{}c@{}}QS-\\ DRC\end{tabular}} &
  \multicolumn{1}{c|}{\textbf{Ours}} &
  \multicolumn{1}{c|}{\textbf{\begin{tabular}[c]{@{}c@{}}Ours\\ NPQC\end{tabular}}} &
  \multicolumn{1}{c|}{\textbf{NGF}} &
  \textbf{\begin{tabular}[c]{@{}c@{}}QS-\\ DRC\end{tabular}} &
  \multicolumn{1}{c|}{\textbf{Ours}} &
  \multicolumn{1}{c|}{\textbf{\begin{tabular}[c]{@{}c@{}}Ours\\ NPQC\end{tabular}}} &
  \multicolumn{1}{c|}{\textbf{NGF}} &
  \textbf{\begin{tabular}[c]{@{}c@{}}QS-\\ DRC\end{tabular}} \\ \hline
\textbf{Armadillo (1.14)} &
  \multicolumn{1}{c|}{\textbf{4.38}} &
  \multicolumn{1}{c|}{6.18} &
  \multicolumn{1}{c|}{5.57} &
  5.20 &
  \multicolumn{1}{c|}{\textbf{3.56}} &
  \multicolumn{1}{c|}{5.28} &
  \multicolumn{1}{c|}{4.12} &
  3.83 &
  \multicolumn{1}{c|}{\textbf{3.07}} &
  \multicolumn{1}{c|}{4.74} &
  \multicolumn{1}{c|}{3.82} &
  3.17 &
  \multicolumn{1}{c|}{2.93} &
  \multicolumn{1}{c|}{4.15} &
  \multicolumn{1}{c|}{3.80} &
  \textbf{2.76} \\
\textbf{Dragon (12.51)} &
  \multicolumn{1}{c|}{\textbf{19.16}} &
  \multicolumn{1}{c|}{21.12} &
  \multicolumn{1}{c|}{19.69} &
  28.29 &
  \multicolumn{1}{c|}{16.38} &
  \multicolumn{1}{c|}{19.84} &
  \multicolumn{1}{c|}{\textbf{16.12}} &
  19.79 &
  \multicolumn{1}{c|}{\textbf{11.54}} &
  \multicolumn{1}{c|}{14.33} &
  \multicolumn{1}{c|}{11.79} &
  12.57 &
  \multicolumn{1}{c|}{\textbf{8.76}} &
  \multicolumn{1}{c|}{11.63} &
  \multicolumn{1}{c|}{10.36} &
  10.16 \\
\textbf{Einstein (8.44)} &
  \multicolumn{1}{c|}{\textbf{7.54}} &
  \multicolumn{1}{c|}{8.66} &
  \multicolumn{1}{c|}{7.89} &
  12.46 &
  \multicolumn{1}{c|}{\textbf{5.55}} &
  \multicolumn{1}{c|}{7.23} &
  \multicolumn{1}{c|}{6.32} &
  6.25 &
  \multicolumn{1}{c|}{\textbf{4.11}} &
  \multicolumn{1}{c|}{6.86} &
  \multicolumn{1}{c|}{4.55} &
  5.01 &
  \multicolumn{1}{c|}{\textbf{3.27}} &
  \multicolumn{1}{c|}{4.89} &
  \multicolumn{1}{c|}{3.98} &
  4.19 \\
\textbf{Ganesha (31.61)} &
  \multicolumn{1}{c|}{\textbf{8.85}} &
  \multicolumn{1}{c|}{10.41} &
  \multicolumn{1}{c|}{9.76} &
  10.76 &
  \multicolumn{1}{c|}{7.66} &
  \multicolumn{1}{c|}{8.54} &
  \multicolumn{1}{c|}{\textbf{7.60}} &
  8.79 &
  \multicolumn{1}{c|}{\textbf{5.14}} &
  \multicolumn{1}{c|}{7.07} &
  \multicolumn{1}{c|}{6.25} &
  7.97 &
  \multicolumn{1}{c|}{\textbf{4.16}} &
  \multicolumn{1}{c|}{6.16} &
  \multicolumn{1}{c|}{5.25} &
  5.29 \\ 
\textbf{Gargoyle (2.32)} &
  \multicolumn{1}{c|}{\textbf{2.54}} &
  \multicolumn{1}{c|}{3.67} &
  \multicolumn{1}{c|}{2.73} &
  5.38 &
  \multicolumn{1}{c|}{2.51} &
  \multicolumn{1}{c|}{3.38} &
  \multicolumn{1}{c|}{\textbf{2.50}} &
  4.17 &
  \multicolumn{1}{c|}{2.47} &
  \multicolumn{1}{c|}{3.12} &
  \multicolumn{1}{c|}{\textbf{2.39}} &
  2.75 &
  \multicolumn{1}{c|}{2.21} &
  \multicolumn{1}{c|}{2.89} &
  \multicolumn{1}{c|}{2.35} &
  \textbf{1.91} \\
\textbf{Gnome (2.25)} &
  \multicolumn{1}{c|}{\textbf{6.02}} &
  \multicolumn{1}{c|}{7.95} &
  \multicolumn{1}{c|}{6.38} &
  10.30 &
  \multicolumn{1}{c|}{\textbf{3.62}} &
  \multicolumn{1}{c|}{5.12} &
  \multicolumn{1}{c|}{4.03} &
  4.10 &
  \multicolumn{1}{c|}{\textbf{2.32}} &
  \multicolumn{1}{c|}{3.77} &
  \multicolumn{1}{c|}{2.88} &
  2.68 &
  \multicolumn{1}{c|}{1.92} &
  \multicolumn{1}{c|}{2.92} &
  \multicolumn{1}{c|}{1.89} &
  \textbf{1.77} \\
\textbf{Head (4.23)} &
  \multicolumn{1}{c|}{\textbf{2.18}} &
  \multicolumn{1}{c|}{3.05} &
  \multicolumn{1}{c|}{2.68} &
  9.78 &
  \multicolumn{1}{c|}{\textbf{1.83}} &
  \multicolumn{1}{c|}{2.40} &
  \multicolumn{1}{c|}{2.19} &
  4.86 &
  \multicolumn{1}{c|}{\textbf{1.40}} &
  \multicolumn{1}{c|}{2.01} &
  \multicolumn{1}{c|}{1.81} &
  1.96 &
  \multicolumn{1}{c|}{1.11} &
  \multicolumn{1}{c|}{1.84} &
  \multicolumn{1}{c|}{1.24} &
  \textbf{0.95} \\
\textbf{Lucy (6.08)} &
  \multicolumn{1}{c|}{\textbf{6.39}} &
  \multicolumn{1}{c|}{8.29} &
  \multicolumn{1}{c|}{8.06} &
  16.65 &
  \multicolumn{1}{c|}{\textbf{4.98}} &
  \multicolumn{1}{c|}{7.07} &
  \multicolumn{1}{c|}{5.96} &
  7.80 &
  \multicolumn{1}{c|}{\textbf{3.64}} &
  \multicolumn{1}{c|}{4.97} &
  \multicolumn{1}{c|}{4.02} &
  4.61 &
  \multicolumn{1}{c|}{\textbf{2.93}} &
  \multicolumn{1}{c|}{4.41} &
  \multicolumn{1}{c|}{3.47} &
  3.70 \\
\textbf{Metatron (1.63)} &
  \multicolumn{1}{c|}{\textbf{3.95}} &
  \multicolumn{1}{c|}{5.60} &
  \multicolumn{1}{c|}{5.26} &
  5.43 &
  \multicolumn{1}{c|}{\textbf{3.10}} &
  \multicolumn{1}{c|}{3.89} &
  \multicolumn{1}{c|}{4.05} &
  3.83 &
  \multicolumn{1}{c|}{\textbf{2.20}} &
  \multicolumn{1}{c|}{2.91} &
  \multicolumn{1}{c|}{3.89} &
  2.57 &
  \multicolumn{1}{c|}{1.97} &
  \multicolumn{1}{c|}{2.67} &
  \multicolumn{1}{c|}{3.46} &
  \textbf{1.84} \\
\textbf{XYZ (2.95)} &
  \multicolumn{1}{c|}{\textbf{7.59}} &
  \multicolumn{1}{c|}{9.71} &
  \multicolumn{1}{c|}{8.97} &
  15.91 &
  \multicolumn{1}{c|}{\textbf{6.16}} &
  \multicolumn{1}{c|}{8.65} &
  \multicolumn{1}{c|}{7.46} &
  7.73 &
  \multicolumn{1}{c|}{\textbf{4.67}} &
  \multicolumn{1}{c|}{6.86} &
  \multicolumn{1}{c|}{5.56} &
  5.87 &
  \multicolumn{1}{c|}{\textbf{4.05}} &
  \multicolumn{1}{c|}{5.85} &
  \multicolumn{1}{c|}{5.13} &
  4.91 \\ \hline
\textbf{Mean Error} &
  \multicolumn{1}{c|}{\textbf{6.86}} &
  \multicolumn{1}{c|}{8.48} &
  \multicolumn{1}{c|}{7.70} &
  12.02 &
  \multicolumn{1}{c|}{\textbf{5.53}} &
  \multicolumn{1}{c|}{7.14} &
  \multicolumn{1}{c|}{6.04} &
  7.12 &
  \multicolumn{1}{c|}{\textbf{4.06}} &
  \multicolumn{1}{c|}{5.66} &
  \multicolumn{1}{c|}{4.70} &
  4.92 &
  \multicolumn{1}{c|}{\textbf{3.33}} &
  \multicolumn{1}{c|}{4.74} &
  \multicolumn{1}{c|}{4.09} &
  3.75 \\ \hline
\end{tabular}%
\caption{$d_{norm}$ evaluated between the reconstructed and original meshes in degrees ($\degree$) is shown for all meshes and methods.}
\label{tab:results-dnorm}
\end{table*}

\begin{figure*}
    \centering
    \includegraphics[width=1.0\linewidth]{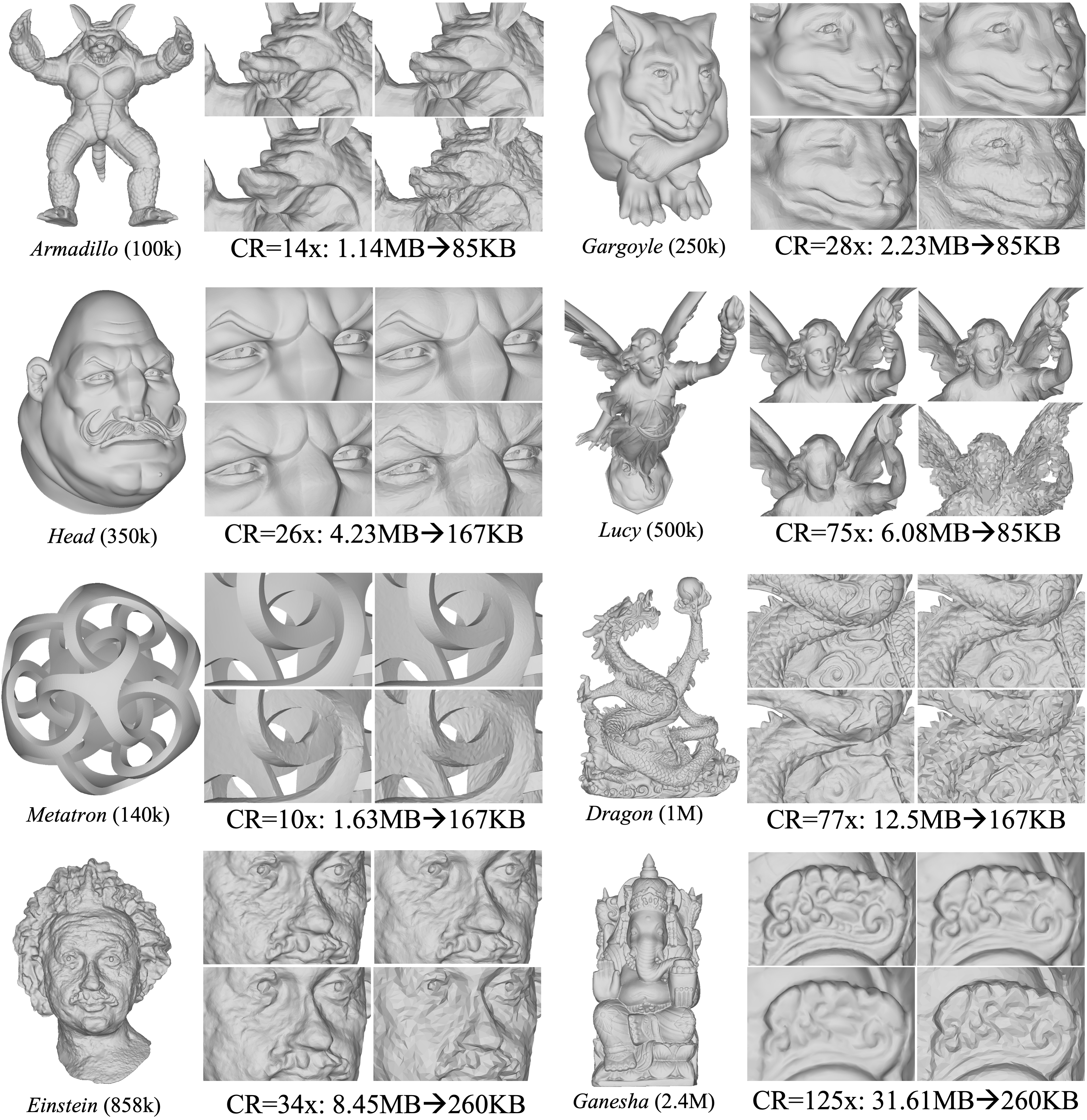}
    \caption{Images of the outputs obtained by compressing and reconstructing various meshes (\# faces in brackets) using different methods are shown for visual comparison. Starting from the top left in a clockwise order we show images corresponding to the original, our method, NGF, and QS-DRC respectively. Our method produces reconstructions with noticeably better visual fidelity compared to the baselines over a wide range of compression ratios by preserving finer details accurately. While NGF also performs admirable compression, it is prone to errors and reconstructing with reduced fidelity in complex geometric regions. When a mesh has high geometric complexity and/or when the compression ratio is high, QS-DRC tends to produce rough-looking surfaces due to the quantization of vertex coordinates and insufficient geometric resolution if the action of QSLIM is prominent.}
    \label{fig:results}
\end{figure*}
\begin{figure*}
    \centering
    \includegraphics[width=1.0\linewidth]{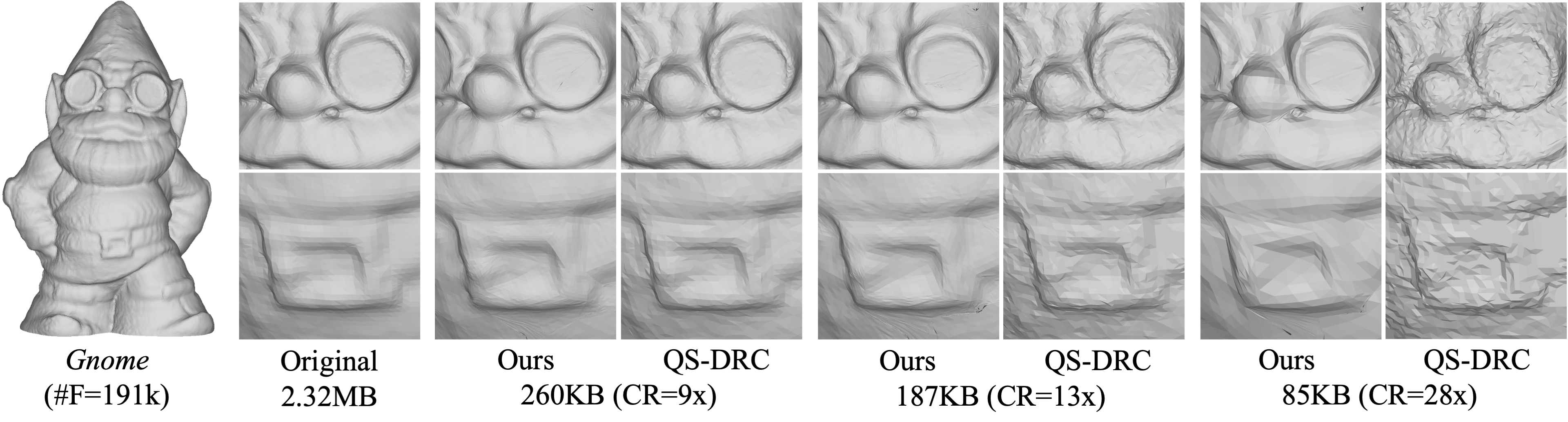}
    \caption{The margin of improved reconstruction quality yielded by our method compared to QS-DRC increases with the compression ratio.
    For a mesh with relatively less geometric complexity such as the one shown here, QS-DRC performs relatively well compared to our method for lower compression ratios. 
    In such cases, QS-DRC could be a more time- and energy-efficient option.}
    \label{fig:simple}
\end{figure*}

We demonstrate the efficacy of our compression algorithm on meshes spanning a diverse range of surface characteristics, genera, and triangle counts.
All meshes are normalized such that the largest dimension of their bounding boxes is one unit.
Our method produces better compression outcomes compared to the baseline methods over a broad range of compression ratios, as shown in Table \ref{tab:results-dpm}, Table \ref{tab:results-dnorm}, and Figure \ref{fig:results}.
Our method outperforms the baselines by much larger extents when the compression ratios are high or the mesh to be compressed has high geometric complexity (see Figure \ref{fig:simple}).
Our choice of baselines includes the following:
\begin{enumerate}
    \item \texttt{NGF} (Neural Geometric Fields) (\cite{ngf}): A method for neural mesh representations and the current state-of-the-art for mesh compression.
    \item \texttt{QS-DRC}: As our method and NGF allow remeshing the meshes during the compression, it is unfair to compare these methods with methods like Draco (\cite{draco}) that encode the vertex connectivity losslessly and use quantization of vertex coordinates. 
    Thus for a fairer comparison, we first simplify meshes to a certain extent using QSLIM (\cite{qslim}) and then compress the simplified connectivity using Draco. 
    We perform a grid search for each mesh to find the target faces for decimation with QSLIM and the number of bits for vertex quantization that minimize the compression error.
    We dub this method as \texttt{QS-DRC}, shorthand for QSLIM-Draco. 
    Using QS-DRC provides a much more competitive baseline compared to using QSLIM and Draco independently.
\end{enumerate}

\begin{figure}
    \centering
    \includegraphics[width=1.0\linewidth]{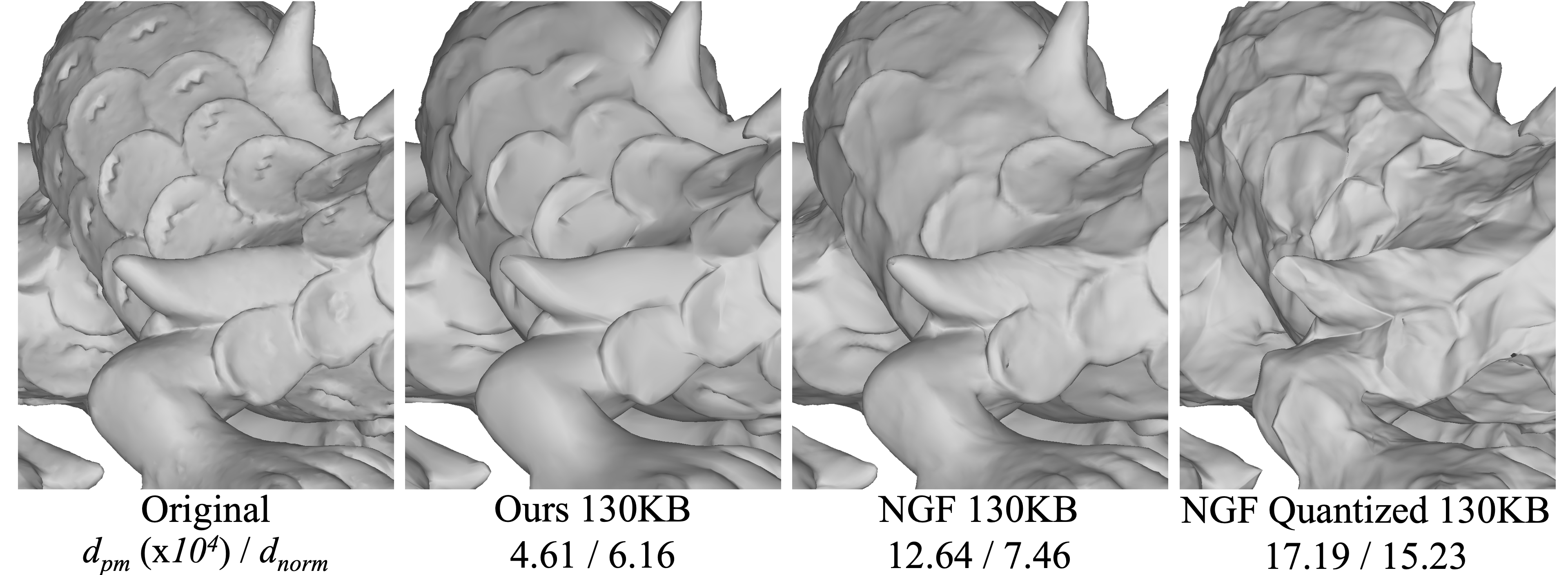}
    \caption{Pruning and quantizing NGF leads to a rough and spiky geometry reconstruction.}
    \label{fig:ngfquant}
\end{figure}

\texttt{NGF} minimizes the rendering loss to reconstruct the geometry of meshes.
This process is sensitive to hyperparameters such as rendering resolution, camera intrinsics, camera distance, and (anti-)aliasing.
On the other hand, our method utilizes a simpler training objective of fitting an MLP to the displacement field generated using SSP, allowing it to achieve better metrics.
However, It is important to note that \texttt{NGF} does not employ pruning, post-training quantization, or entropy coding of parameters.
Thus for a fair comparison, we also included the performance of our method without applying the post-training model compression pipeline in the "Ours NPQC" (No Pruning Quantization and Coding) columns of Tables \ref{tab:results-dpm} and \ref{tab:results-dnorm}. 
Ours NPQC outperformed \texttt{NGF} in terms of $d_{pm}$.
On the other hand, Ours NPQC lags \texttt{NGF} in terms of $d_{norm}$, which may be due to the explicit minimization of normal error in the pipeline of \texttt{NGF}.
Conversely, applying similar post-training compression to NGF does not yield better results.
This is because \texttt{NGF} consists of an MLP with just two hidden layers, which only consumes a small fraction of the overall memory, while most of the used memory is occupied by per-vertex features. 
When vertex-wise features are pruned and quantized, their performance deteriorates severely as shown in Figure \ref{fig:ngfquant}.

\begin{figure}
    \centering
    \includegraphics[width=1.0\linewidth]{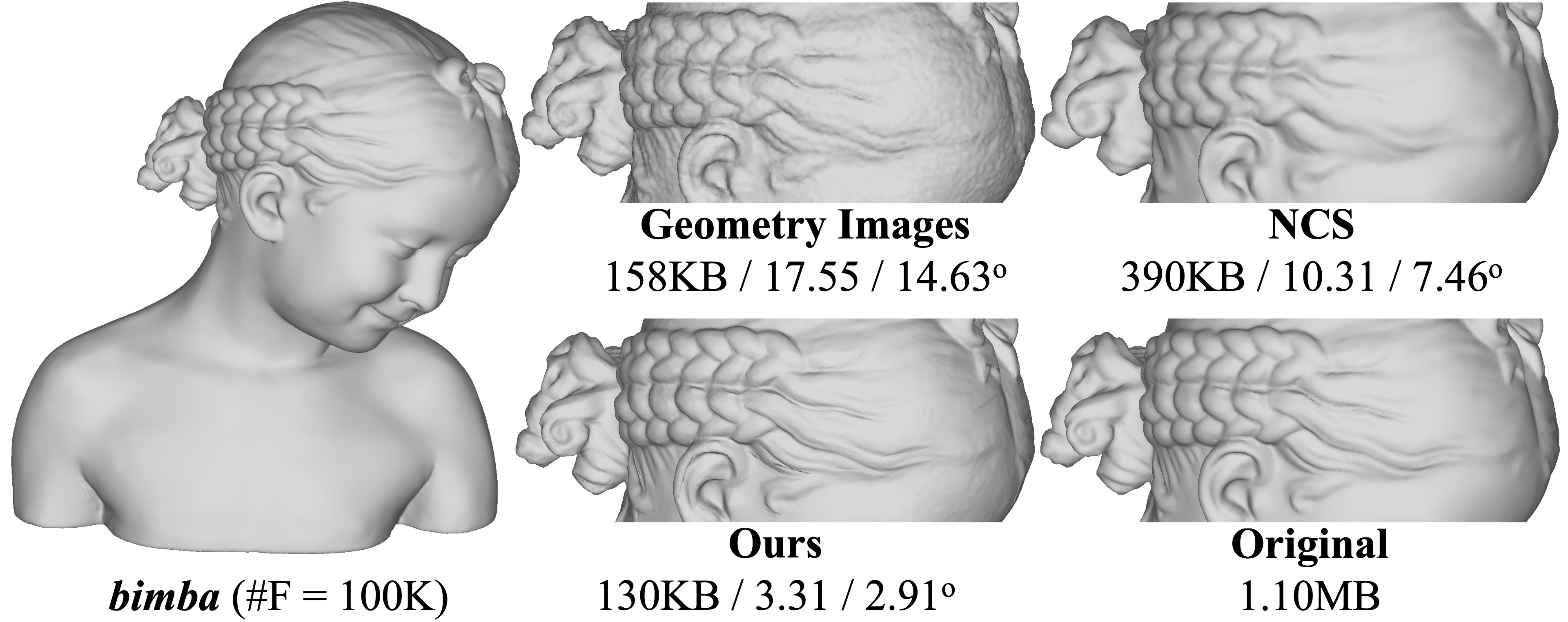}
    \caption{Geometry Images \cite{geoimg} shows quantization-like artifacts and NCS \cite{morreale2022neural} fails to reconstruct finer details while our method retains the geometry more faithfully. 
    The size of the compressed representation, $d_{pm}(\times10^4)$ and $d_{norm}$ are displayed for each method}
    \label{fig:imgmaps}
\end{figure}

We also compared our method against Geometry Images \cite{geoimg} and Neural Convolutional Surfaces (NCS)\cite{morreale2022neural} as shown in Figure \ref{fig:imgmaps}.
These methods involve encoding position or displacement maps of mesh surfaces in the two-dimensional UV plane, using standard image compression techniques and are thus broadly related to our work.
Our method performs much better than those owing to the higher compressibility of INRs and the more accurate ground truth displacements predicted by SSP.

\begin{figure}
    \centering
    \includegraphics[width=1.0\linewidth]{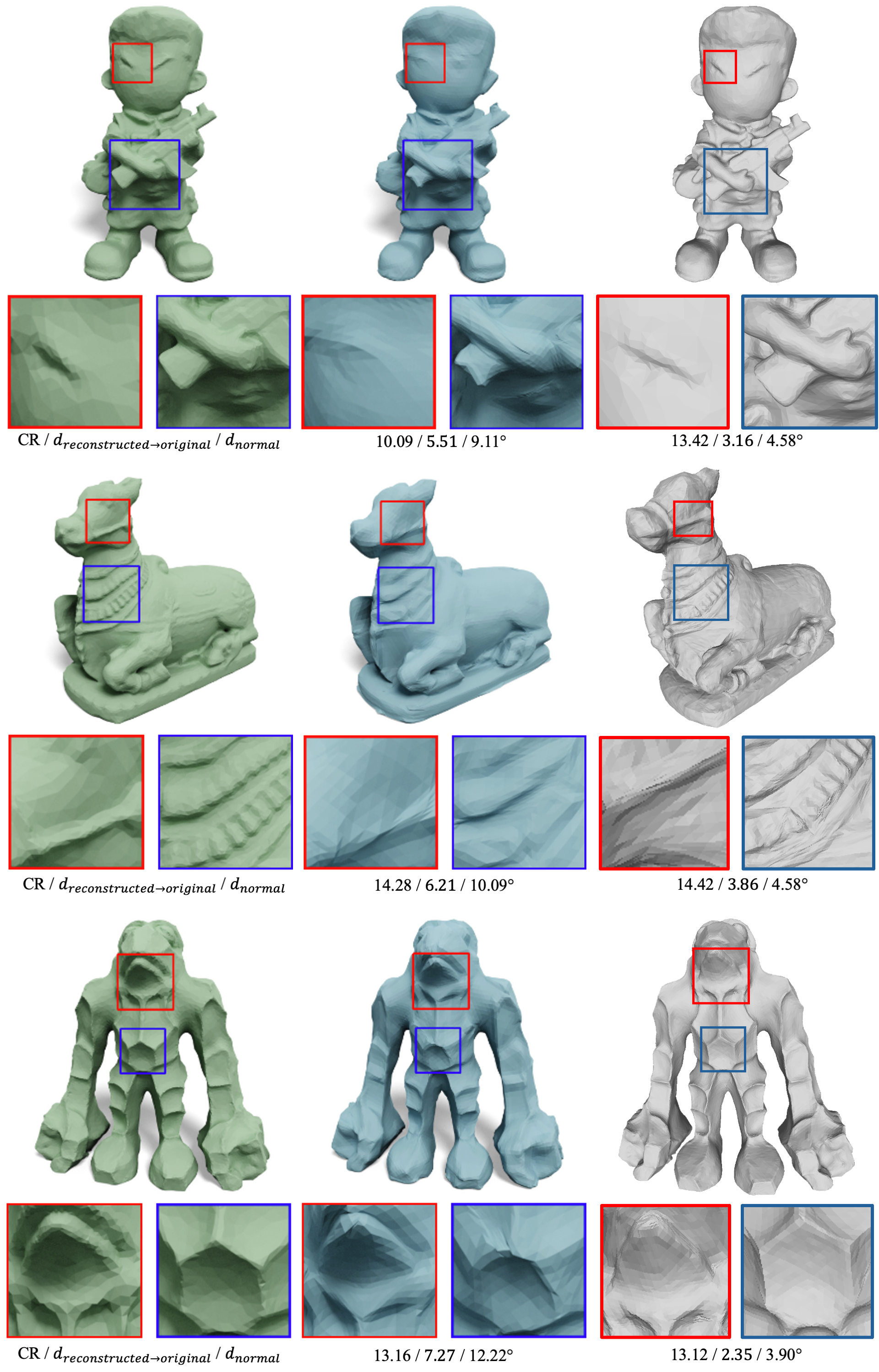}
    \caption{Comparison with Neural Progressive Meshes (NPM). 
    The mesh in the left column (green) is the original, the middle column (blue) is NPM, and the right (gray) is ours. The point-to-mesh errors are scaled $\times10^4$}
    \label{fig:npm}
\end{figure}

Additionally, we also compare our method with Neural Progressive Meshes (NPM) (\cite{npm}) in Figure \ref{fig:npm}. 
With its code, trained model, and training dataset not being available publicly, we present comparisons only with a few of the meshes whose images are available in their paper and were easily found on the internet.

The size of original meshes reported in Tables \ref{tab:results-dpm} and \ref{tab:results-dnorm}  is equal to $32\times3v + 3f\log_2 v$ bits, where $v$ and $f$ are the number of vertices and faces in the mesh respectively.
The size of the coarse meshes which are a part of the compressed representation are also governed by the same formula.
In practice, the coarse meshes included in our compressed representation can be represented much more compactly by applying any of the lossless mesh compression methods (such as the ones mentioned in Section \ref{sec:related}) to yield even higher compression ratios without any reduction of quality.

To evaluate the quality of a surface's reconstruction, we measure the mean point to mesh distance ($d_{pm}$) and the mean normal error ($d_{norm}$). 
To calculate $d_{pm}$ between meshes $\text{M}_a$ and $\text{M}_b$, we first sample 1 million points on the surface of $\text{M}_a$, then for each of these points we find the closest point on the surface of $\text{M}_b$ and measure the average distance between them to obtain $d_{a\rightarrow b}$. 
Similarly, we also evaluate $d_{b\rightarrow a}$. 
Finally, $d_{pm}(\text{M}_\text{a}, \text{M}_\text{b}) = d_{a\rightarrow b}+d_{a\rightarrow b}$.
In a similar fashion, $d_{norm}$ is calculated to be the average difference between the angles of the normal vectors (in degrees) of the same set of points used during the evaluation of $d_{pm}$.

These metrics for evaluation are different from those used by authors of recent works. 
\cite{ngf} measure the Chamfer distance between the vertices of $\text{M}_a$ and $\text{M}_b$.
This method is sensitive to vertex resolution, which is not truly representative of the geometric information a mesh depicts.
Measuring Chamfer distance between the sets of vertices in the meshes also fails to account for the continuous surface manifold.
On the other hand, \cite{npm} only report $d_{reconstructed\rightarrow original}$, which does not account for directional asymmetry.

\setlength{\tabcolsep}{3.1pt}
\renewcommand{\arraystretch}{1.4}
\begin{table*}[t]
\customsize
\centering
\begin{tabular}{|c|cccc|cccc|cccc|}
\hline
\multirow{2}{*}{\textbf{\begin{tabular}[c]{@{}c@{}}\\ $\text{V}_\text{coarse}$ \\\end{tabular}}} &
  \multicolumn{4}{c|}{\textbf{2 subdivisions}} &
  \multicolumn{4}{c|}{\textbf{3 subdivisions}} &
  \multicolumn{4}{c|}{\textbf{4 subdivisions}} \\ \cline{2-13} 
 &
  \multicolumn{1}{c|}{\textbf{SSP GT}} &
  \multicolumn{1}{c|}{\textbf{\begin{tabular}[c]{@{}c@{}}Without \\ Pruning\end{tabular}}} &
  \multicolumn{1}{c|}{\textbf{\begin{tabular}[c]{@{}c@{}}With \\ Pruning\end{tabular}}} &
  \textbf{\begin{tabular}[c]{@{}c@{}}After \\ Quantizing\end{tabular}} &
  \multicolumn{1}{c|}{\textbf{SSP GT}} &
  \multicolumn{1}{c|}{\textbf{\begin{tabular}[c]{@{}c@{}}Without \\ Pruning\end{tabular}}} &
  \multicolumn{1}{c|}{\textbf{\begin{tabular}[c]{@{}c@{}}With \\ Pruning\end{tabular}}} &
  \textbf{\begin{tabular}[c]{@{}c@{}}After \\ Quantizing\end{tabular}} &
  \multicolumn{1}{c|}{\textbf{SSP GT}} &
  \multicolumn{1}{c|}{\textbf{\begin{tabular}[c]{@{}c@{}}Without \\ Pruning\end{tabular}}} &
  \multicolumn{1}{c|}{\textbf{\begin{tabular}[c]{@{}c@{}}With \\ Pruning\end{tabular}}} &
  \textbf{\begin{tabular}[c]{@{}c@{}}After \\ Quantizing\end{tabular}} \\ \hline
\textbf{500} &
  \multicolumn{1}{c|}{28.64/30.24} &
  \multicolumn{1}{c|}{29.23/30.24} &
  \multicolumn{1}{c|}{29.23/30.25} &
  31.01/30.43 &
  \multicolumn{1}{c|}{15.57/26.25} &
  \multicolumn{1}{c|}{16.15/26.65} &
  \multicolumn{1}{c|}{16.42/26.67} &
  18.48/27.20 &
  \multicolumn{1}{c|}{8.21/22.42} &
  \multicolumn{1}{c|}{8.79/22.86} &
  \multicolumn{1}{c|}{8.88/22.89} &
  9.72/23.41 \\ \hline
\textbf{1000} &
  \multicolumn{1}{c|}{22.46/26.25} &
  \multicolumn{1}{c|}{23.51/26.28} &
  \multicolumn{1}{c|}{23.55/26.30} &
  24.01/26.48 &
  \multicolumn{1}{c|}{12.11/21.69} &
  \multicolumn{1}{c|}{13.27/21.98} &
  \multicolumn{1}{c|}{13.35/22.02} &
  14.24/22.23 &
  \multicolumn{1}{c|}{6.15/17.03} &
  \multicolumn{1}{c|}{6.59/17.27} &
  \multicolumn{1}{c|}{6.84/17.47} &
  7.55/18.40 \\ \hline
\textbf{2000} &
  \multicolumn{1}{c|}{9.31/17.58} &
  \multicolumn{1}{c|}{9.89/17.64} &
  \multicolumn{1}{c|}{9.93/17.64} &
  10.63/17.65 &
  \multicolumn{1}{c|}{4.42/11.54} &
  \multicolumn{1}{c|}{6.24/11.67} &
  \multicolumn{1}{c|}{6.29/12.04} &
  7.59/12.62 &
  \multicolumn{1}{c|}{2.00/6.33} &
  \multicolumn{1}{c|}{5.84/8.03} &
  \multicolumn{1}{c|}{6.11/8.41} &
  6.81/8.98 \\ \hline
\textbf{4000} &
  \multicolumn{1}{c|}{4.91/11.86} &
  \multicolumn{1}{c|}{5.56/11.88} &
  \multicolumn{1}{c|}{5.95/12.03} &
  6.52/12.14 &
  \multicolumn{1}{c|}{2.22/6.37} &
  \multicolumn{1}{c|}{5.63/8.04} &
  \multicolumn{1}{c|}{5.95/8.28} &
  6.46/8.76 &
  \multicolumn{1}{c|}{0.96/3.07} &
  \multicolumn{1}{c|}{5.42/7.55} &
  \multicolumn{1}{c|}{5.81/7.97} &
  6.22/8.83 \\ \hline
\textbf{8000} &
  \multicolumn{1}{c|}{2.68/7.72} &
  \multicolumn{1}{c|}{4.80/8.18} &
  \multicolumn{1}{c|}{5.09/8.58} &
  5.64/8.97 &
  \multicolumn{1}{c|}{1.24/3.64} &
  \multicolumn{1}{c|}{5.16/7.09} &
  \multicolumn{1}{c|}{5.49/7.78} &
  6.01/8.51 &
  \multicolumn{1}{c|}{0.63/2.49} &
  \multicolumn{1}{c|}{4.93/6.79} &
  \multicolumn{1}{c|}{5.33/7.61} &
  5.95/8.23 \\ \hline
\end{tabular}%
\caption{Effect of the size of the coarse mesh and the number of subdivisions on the reconstruction quality when the hyperparameters of the INR model are fixed at $l=32, k=96$. 
\textbf{SSP GT} shows the reconstruction quality of the re-meshed approximation of the \textbf{dragon} mesh used as ground truth for training the INR. 
The reconstruction quality before and after pruning and quantization are also shown. 
Each cell shows $d_{pm}$ and $d_{norm}$ (in degrees) for the corresponding configuration.}
\label{tab:ablation}
\end{table*}

\subsection{Experiment Setup}
\label{subsec:setup}

\begin{table}
\resizebox{\columnwidth}{!}{%
\centering
\begin{tabular}{|c|c|c|c|c|}
\hline
  \textbf{\begin{tabular}[c]{@{}c@{}}Compressed \\ Size\end{tabular}} &
  \textbf{\begin{tabular}[c]{@{}c@{}}Vertices in \\ coarse mesh \\ ($=\text{V}_\text{coarse}$)\end{tabular}} &
  \textbf{\begin{tabular}[c]{@{}c@{}}Number of \\ sub-divisions \\ ($=s$)\end{tabular}} &
  \textbf{\begin{tabular}[c]{@{}c@{}}Number of \\ hidden layers \\ ($=l$)\end{tabular}} &
  \textbf{\begin{tabular}[c]{@{}c@{}}Hidden layer \\ size ($=k$)\end{tabular}} \\ \hline
85KB  & 2000 & 2 & 20 & 56 \\ \hline
130KB & 2500 & 2 & 24 & 70 \\ \hline
187KB & 3000 & 3 & 28 & 82 \\ \hline
260KB & 3500 & 3 & 32 & 96 \\ \hline
\end{tabular}%
}
\caption{Hyper-parameters for different configurations that were used to compress meshes to different bitrates yielding the results in table 1.}
\label{tab:config}
\end{table}

We used our method to compress all meshes to 4 bitrates - 85, 130, 187, and 260 kilobytes(KB).
The hyperparameters of our method to obtain these bitrates are shown in table \ref{tab:config}.
The number of layers with one-ring feature accumulation, $g=4$, was the same for all configurations.
We also fixed $Q$, the dimensionality of the positional embedding space to be 10 in all configurations.
In all cases, the network was optimized using the AdamW optimizer \cite{adamw} to minimize the mean squared error between the displacements predicted by the MLP, and the ground truth vectors computed using SSP. 
The optimization was performed over 3500 epochs with a batch size of 2048, and an initial learning rate of $1e-3$ with a cosine decay scheduler. 
We obtained an overall pruning ratio of 50\% through $z=5$ progressive pruning iterations, with an 87\% pruning ratio introduced in each step.
The pruning was performed after epoch numbers 200, 400, 600, 800, and 1000.
Please visit the \href{https://karthikey97.github.io/nmc/}{project website} for the complete implementation.
\subsection{Ablation Studies}
\label{subsec:ablation}
\begin{figure}
    \centering
    \includegraphics[width=0.48\textwidth]{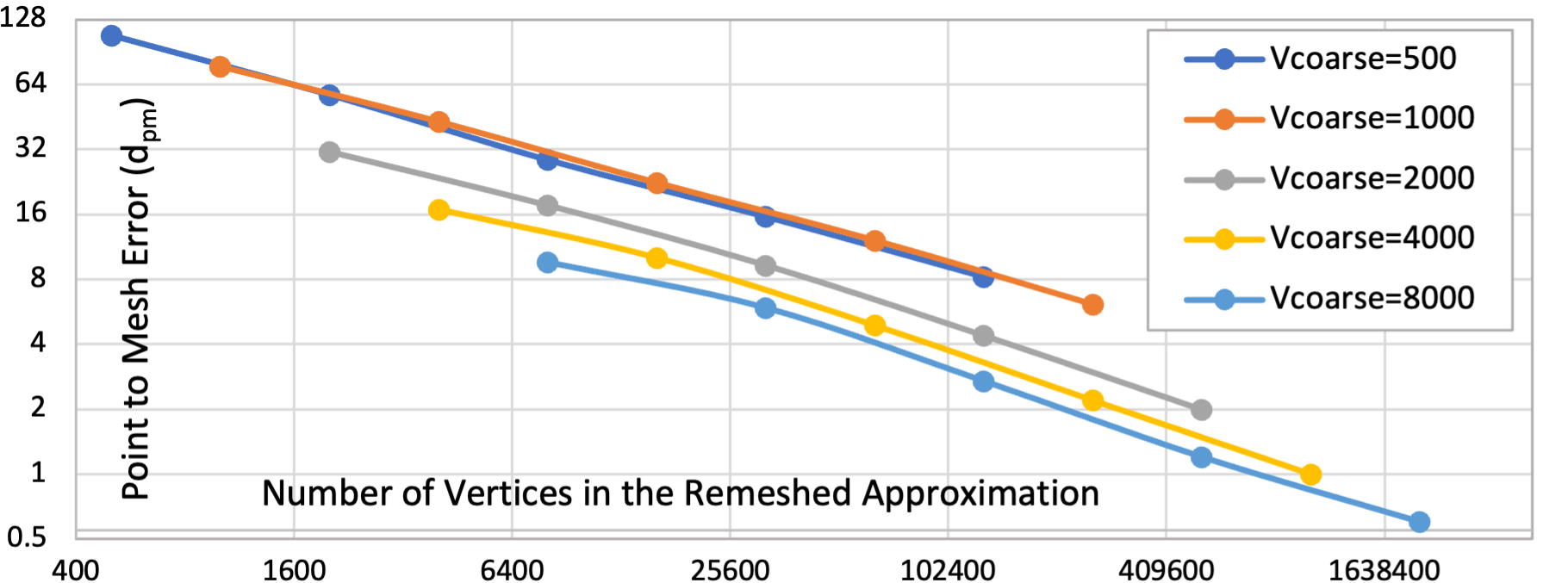}
    \caption{\textbf{Dragon} mesh is first decimated to different resolutions ($\text{V}_\text{coarse}$ and subdivided 0-4 times reaching different vertex resolutions of the remeshed approximations. 
    Using a larger coarse mesh provides a better remeshed version with lower $d_{pm}$ than having more subdivisions. 
    The $d_{pm}$ axis is scaled $\times10^4$.}
    \label{fig:sspabla}
\end{figure}

\begin{table}
    \resizebox{\columnwidth}{!}{
    \centering
    \begin{tabular}{|c|c|c|c|c|c|c|}
        \hline
        \textbf{Prune+EC} & \textbf{Quantize} & \textbf{l=20,k=56} & \textbf{l=24,k=70} & \textbf{l=28,k=82} & \textbf{l=32,k=96} \\
        \hline
        \textbf{No} & \textbf{No} & 300KB & 536KB & 836KB & 1284KB \\
        \hline
        \textbf{No} & \textbf{Yes} & 75KB & 134KB & 209KB & 321KB \\
        \hline
        \textbf{Yes} & \textbf{No} & 226KB & 453KB & 685KB & 804KB \\
        \hline
        \textbf{Yes} & \textbf{Yes} & 45KB & 79KB & 102KB & 160KB \\
        \hline
    \end{tabular}
    }
    \caption{Variation in storage size of MLP with pruning+entropy coding and quantization}
    \label{tab:memabla}
\end{table}

As explained in section \ref{subsec:datasest}, performing a finite number of subdivisions of a coarse mesh will only lead to an approximate reconstruction of the original mesh surface.
Using this approximation as the dataset for training the INR introduces bounds on reconstruction quality.
This is because the sampling error cannot be recovered even if the INR encodes the sampled displacement field perfectly.
The \textbf{SSP GT} columns of table \ref{tab:ablation} show the reconstruction accuracy of the re-meshed approximations generated using SSP which were used as the ground truth when training the neural representations.
These act as an upper bound on the reconstruction quality achievable by the INR for a given configuration of coarse mesh size and number of subdivisions.
The plot in Figure \ref{fig:sspabla} shows that remeshing quality is better when larger coarse meshes are used compared to performing more subdivisions while the number of vertices in the remeshed approximation stays the same. 
However, unlike increasing the number of vertices in the coarse mesh, performing more subdivisions incurs no additional increase in the size of the compressed representation.
Table \ref{tab:ablation} also shows the drop in reconstruction quality with the application of pruning and quantization while the size of the INR is fixed at $l=32, k=96$.
It may be noted that pruning has minimal or no effect when the size of the INR used to encode the sampled displacement field has adequate learning capacity. 
While in such cases, the reconstruction quality of the remeshed ground truth (SSP GT) is the bottleneck, when the sampling rate of the ground truth displacement field increases (such as when $\text{V}_\text{coarse}$ and number of subdivision $s$ are high), the size of the INR model becomes the bottleneck with the reconstruction quality of the INR being much higher than the quality of the remeshed ground truth.

Performing model quantization always incurs a reconstruction quality cost of approximately $+1.2$ points increase in $d_{pm}$ and approximately $+0.5\degree$ increase in $d_{norm}$.
The reduction of memory due to quantization and pruning+entropy coding is shown in table \ref{tab:memabla}.

Not using the one-ring feature accumulation reduces the $d_{pm}$ and $d_{norm}$ by factors of 15\% and 4\%, respectively on average. 
Using full graph convolutions improves $d_{pm}$ by a factor of only 3\% while increasing the training time by around 70\% and the GPU memory usage by 300\%.

\subsection{Runtime and Memory Usage}
\begin{table}
    \resizebox{\columnwidth}{!}{
    \centering
    \begin{tabular}{@{}|c|c|c|c|@{}}
        \hline
        \textbf{Method} & \textbf{Encoding Time} & \textbf{Decoding Time} & \textbf{GPU Memory} \\
        \hline
        \textbf{Ours (s=2)}        & 17-22 mins       & 270 ms        & 1.7 GB    \\
        \textbf{Ours (s=3)}        & 65-70 mins       & 800 ms        & 2.5 GB    \\
        \textbf{NGF ($<$140KB)}      & 8-9 mins        & 200 ms        & 8 GB       \\
        \textbf{NGF ($>$140KB)}      & 25-28 mins       & 750 ms        & 8 GB       \\
        \textbf{QS-DRC}            & 3 mins        & 10 ms         & None       \\
        \hline
    \end{tabular}
    }
    \caption{Encoding and decoding times with GPU Memory Usage for various methods for the \textit{Ganesha} mesh with 2.1M triangles. Note that the time for encoding using QS-DRC includes the time taken for performing the grid search which involves multiple decimation operations.}
    \label{tab:runtime}
\end{table}

Table \ref{tab:runtime} shows the runtime and memory usage of our method and the previous SOTA baseline when run on Nvidia GTX 1080Ti GPU for the Ganesha mesh with 2.1M triangles.
The main hyperparameter that affects the runtime is the number of times the coarse mesh is subdivided ($s$).
Increasing the $s$ by one increases the number of samples in the training dataset four times, increasing the training time proportionally, while changing the size of the INR does not affect the training time drastically.
The training dataset is generated only once per training at the beginning and involves subdividing the coarse mesh a fixed number of times and applying SSP for all vertices on the subdivided mesh.
This process takes less than a minute.
While providing significant improvement in compression quality, our method can compress meshes in the same order of magnitude of time and memory.
The decoding speed of our method is practically instantaneous and similar to other methods that rely on per-mesh optimization.
To improve the speed of encoding of our method, the number of epochs used for fitting the network can be reduced albeit at the cost of quality.
We demonstrate these results in Table \ref{tab:timeabla}
While the encoding times are high for methods that rely on mesh-wise neural network optimization rendering them impractical for real-time and low-latency applications, they are much more suitable for distributing graphical assets to a broad audience, such as the distribution of video games, given the low decoding times.

\begin{table}[]
\centering
\footnotesize
\begin{tabular}{|c|c|c|c|c|c|}
\hline
\textbf{Epochs} & \textbf{1000} & \textbf{2000} & \textbf{3000} & \textbf{4000} & \textbf{5000} \\ \hline
Train Time (mins) & 6   & 12  & 17  & 24  & 30  \\ \hline
\textbf{$d_{pm} (x10^4)$} & 5.89           & 5.55           & 5.19           & 5.13           & 5.08           \\ \hline
\textbf{$d_{normal}$}         & 7.07 & 6.79 & 6.47 & 6.36 & 6.34 \\ \hline
\end{tabular}%
\caption{Effect of varying the number of epochs on the encoding quality for the Lucy model.}
\label{tab:timeabla}
\vspace{-10px}
\end{table}

\subsection{Failure Cases}
\begin{figure}
    \centering
    \includegraphics[width=1.0\linewidth]{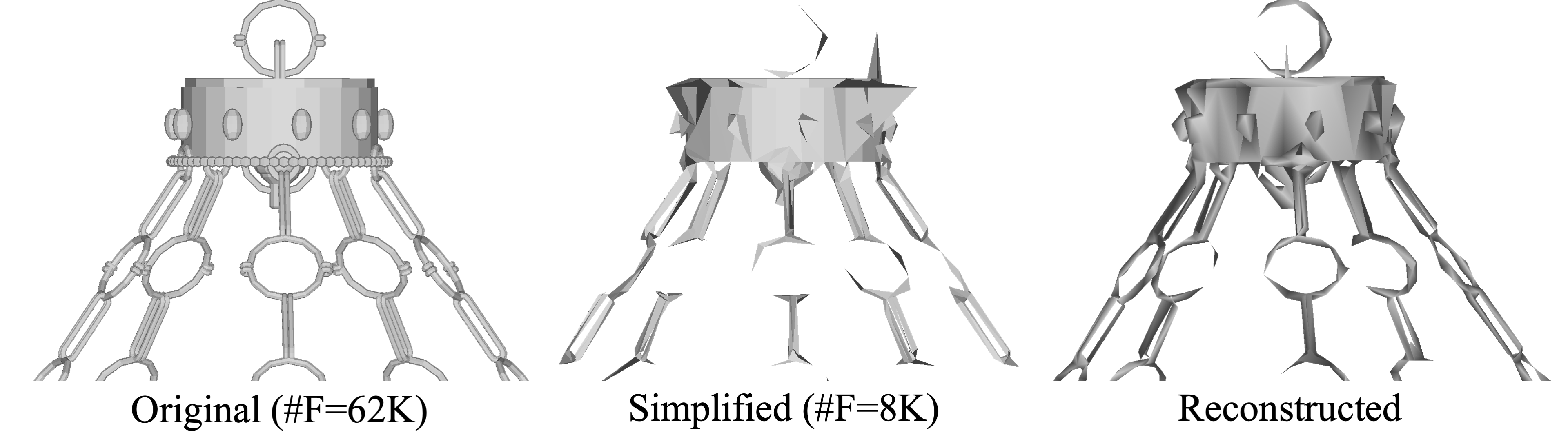}
    \caption{The displacement fails to recover the geometry sufficiently after meshes with thin geometric structures undergo severe simplification.}
    \label{fig:failcase}
\end{figure}

The first step of our method involves building a coarse approximation of the surface to be compressed by simplifying the original mesh and reducing the number of triangles in it drastically (at least $50\times$).
When simplified to a large factor, some meshes, such as the one shown in Figure \ref{fig:failcase}, suffer from broken connectivity and structure.
In such cases, the displacement field is insufficient to recover the geometric structure.
Other methods that also rely on neural overfitting, such as \cite{ngf, jiang2024cofie, npm}, also suffer from the same limitations.

Our method also proves to be redundant when the mesh to be encoded is too simple and contains few geometrically detailed elements. 
In such cases, the displacement field would not significantly improve the reconstruction quality of the simplified surface and thus be redundant. 
For these meshes, QS-DRC would be more efficient due to its much faster encoding process and the loss in quality due to simplification and vertex quantization being negligible.

\section{Conclusion and Scope for Future Research}
\label{sec:conclusion}
This paper presents a state-of-the-art method for compressing 3D meshes that obtains up to $3\times$ less $d_{pm}$ than baselines for the same compression ratios.
The margin of increase in the performance of our method over the existing method increases with the compression ratio allowing generation of more compact encoding of 3D triangle meshes.
Despite the exceptional performance, our method has some current limitations that may motivate further research on this topic:
\begin{enumerate}
    \item While per-mesh optimization methods like ours, \cite{ngf}, and \cite{nglod} can achieve impressive compression ratios, there is still a need to find ways to further accelerate the encoding speed of these methods to improve their practicality for more use cases.
    \item While our method can be easily modified to compress UV parameters of vertices by performing mesh simplification in a manner that is UV-parameterization aware, there is also scope for extending the algorithm to compress the attribute maps as well by incorporating techniques proposed by \cite{geoscaler} or \cite{texbake}.
    \item As SSP relies on flattening mesh patches onto a 2D plane, it requires the surface to be edge-manifold.
    Because our method relies on SSP to build a training dataset, it has the same limitation. 
    If the non-manifold aspect of the mesh needs to be preserved, as a workaround, the mesh could be partitioned by splitting the mesh along such edges and compressing the components independently.
    Otherwise, the surface could be remeshed to obtain a triangulation amenable to compression with our algorithm.
\end{enumerate}

\noindent 
\textbf{Acknowledgements} 
\newline The authors thank the Texas Advance Computing Center and the National Science Foundation AI Institute for Foundations of Machine Learning (Grant 2019844) for providing compute resources that contributed to our research.
\bibliographystyle{eg-alpha-doi} 
\bibliography{egbibsample}       


\end{document}